\documentclass[english,floatfix,twocolumn,showkeys,superscriptaddress,nofootinbib,aps,prd]{revtex4}
\usepackage[latin1]{inputenc}
\usepackage[T1]{fontenc}
\usepackage{lmodern}
\usepackage{amsmath}
\usepackage{amssymb}
\usepackage{graphicx}
\usepackage{esint}
\usepackage{longtable}
\usepackage{dcolumn}
\usepackage{babel} 
\usepackage{csquotes} 
\usepackage{color} 

\begin{document}

\title{Wormholes from beyond}

\author{Juliano C. S. Neves} 
\email{juliano.c.s.neves@gmail.com}
\affiliation{Instituto de Ciência e Tecnologia, Universidade Federal de Alfenas, \\ Rodovia José Aurélio Vilela,
11999, CEP 37715-400 Poços de Caldas, MG, Brazil}

\begin{abstract}
In a brane-world context in which our universe would be a four-dimensional brane embedded into a
five-dimensional spacetime or bulk, wormhole geometries are induced on branes. 
In this article, the Morris-Thorne wormhole and the Molina-Neves wormhole are obtained on the
brane using the Nakas-Kanti approach, which starts from a regular five-dimensional spacetime to
obtain known black hole and wormhole solutions on the four-dimensional brane. From the bulk 
perspective, these wormholes are five-dimensional solutions supported by an exotic fluid, but from the
brane perspective, such objects are wormholes not supported by any fields or particles that live on the 
four-dimensional spacetime. Thus, the cause of these wormholes is the bulk influence on the brane.
\end{abstract}


\maketitle

\section{Introduction}

From brane-world scenarios,  the Randall-Sundrum model I  \cite{RSI} is an interesting attempt to 
solve the hierarchy problem in particle physics with the aid of a finite extra dimension and two 
four-dimensional branes, where one of them would be our universe. The five-dimensional
spacetime in which the two branes are embedded is the anti-de Sitter spacetime, and its
metric presents a warp factor that confines the gravitational interaction close to the four-dimensional branes 
fixing then the hierarchy problem. Later, the Randall-Sundrum II 
 \cite{RSII} develops such an idea assuming just one brane and an infinite extra dimension. The Randall-Sundrum II
 is more interesting for the study of the gravitational phenomenon in this context of extra dimension (see
  Refs. \cite{Maartens,Clifton} for good reviews on the brane-world context). 
 
As for the gravitational interaction, after the seminal article of Chamblin, Hawking and Reall \textit{et al.} \cite{Hawking} 
about the so-called black string, a large number of articles discussed gravitation and cosmology in scenarios 
inspired by the Randall-Sundrum II model. It is common divide these works into two types:
the first one builds spacetime solutions from the four-dimensional brane ignoring
the five-dimensional or bulk solution (see Refs. \cite{Casadio:2001jg,Bronnikov:2003gx,Molina:2010yu,MN,Neves:2012it,Lobo:2007qi,Barcelo:2000ta,Dadhich:2000am,Aliev:2005bi,Neves:2015vga,Neves:2015zia}).
In this strategy, it is argued that the Campbell-Magaard theorems \cite{Seahra:2003eb} 
guarantee, at least locally, the solution in the bulk.
The second type of works builds the bulk spacetime and then \enquote{finds}
 the brane spacetime geometry \cite{Creek:2006je}.
The latter type, for example, is developed by Nakas and Kanti \cite{Nakas,Nakas:2021srr} and is adopted in this article. 

In the context of the Randall-Sundrum II model, the Nakas-Kanti approach
 \cite{Nakas} starts from
the bulk geometry and then a known solution in general relativity (or other context) 
is obtained on the brane. This procedure is equivalent
to  localize the brane world in this scenario of extra dimension. 
The virtue of the Nakas-Kanti approach is to show that the induced metric on the brane 
could be the Schwarzschild solution \cite{Nakas}, the Reissner-Nordström \cite{Nakas:2021srr} or
a regular black hole solutions \cite{Neves:2021dqx,Crispim:2024nou}. Different spacetime solutions
are generated even without any matter fields, particles or sources that live exclusively on the brane. 
The reason for this is that the bulk geometry or its exotic fluid acts on the brane 
inducing known four-dimensional black hole or wormhole solutions. 
Another virtue of the Nakas-Kanti approach is that---contrary to the black string---the 
bulk spacetime is regular. On the other hand,
as mentioned, the approach give us regular or even singular black hole solutions on the brane.
Here I study wormholes solutions in this approach, showing then that the Morris-Thorne \cite{MT}
wormhole could be an induced spacetime on the brane in this context. Also,
I point out that the Molina-Neves wormhole \cite{MN}, an asymptotically de Sitter wormhole, 
obtained in the Randall-Sundrum context 
from the brane without knowing the bulk metric, could also be an induced geometry on the brane using the
Nakas-Kanti approach.  

How can we assert that any four-dimensional metric is an induced metric on the brane?
The answer is the effective four-dimensional field equations on the brane, deduced by 
Shiromizu, Maeda and Sasaki \cite{Shiromizu}. The effective field equations depend on the
bulk energy-momentum tensor. In order to provide known solutions on the brane,
the Nakas-Kanti approach leads to the bulk energy-momentum tensor which generates the desirable 
solution on the brane with the aid of the effective four-dimensional field equations. 

As mentioned, this article is an attempt at finding the Morris-Thorne and Molina-Neves wormholes
on the brane from a regular bulk spacetime. Since the famous article of Morris and Thorne \cite{MT},
in which wormholes are conceived of as both spacetime shortcuts and time machines, the
interest on these objects has increased in the last decades. Wormhole solutions have been 
calculated in several contexts, in the general relativity realm \cite{Visser:1989kh,Lemos:2003jb,Simpson:2018tsi}
 and  in theories and models beyond general relativity, 
like brane worlds 
\cite{Casadio:2001jg,Bronnikov:2003gx,Molina:2010yu,MN,Lobo:2007qi,Barcelo:2000ta} and other
modified theories of gravity \cite{Lobo:2009ip,Harko:2013yb,Kanti:2011jz,Agnese:1995kd,Moraes:2017mir,Ovgun:2018xys,Nicolini:2009gw,Kanti:2011yv,Santos:2023zrj,Frizo:2022jyz}.
In terms of astrophysical observations, there is some effort to interpret recent data using the
wormhole hypothesis instead of the black hole hypothesis \cite{Cardoso:2019rvt,Bambi:2021qfo,Vagnozzi:2022moj,Dai:2019mse,DeFalco:2020afv,Simonetti:2020ivl}. 
Even without any confirmation from observations,
wormhole solutions are still alternative options to interpret astrophysical compact objects.   
 
This article is structured as follows: in Sec. \ref{Sec-II} one presents the Nakas-Kanti approach or how to start
from the bulk spacetime to obtain the induced metric on the brane. 
In Sec. \ref{Sec-III} the brane perspective is described, and the necessary effective four-dimensional field
equations are inserted into the approach. In Sec. \ref{Sec-IV} two wormhole solutions, the Morris-Thorne and
the Molina-Neves wormholes, are successfully induced on the brane, showing then that the Nakas-Kanti approach again works well.
 The final remarks are given in Sec. \ref{Sec-V}. 

In this work, geometrized units are adopted, i.e., $G=c=1$ throughout this article. Capital Latin index runs 
from 0 to 4, and Greek index runs from 0 to 3. 
 
\section{The bulk equations}
\label{Sec-II}
In this section, it is shown how 
the five-dimensional bulk geometry is achieved from the Nakas-Kanti approach \cite{Nakas}. 
We begin with the original five-dimensional spacetime of the 
Randall-Sundrum model II \cite{RSII}, which is given by
\begin{equation}
ds^2=e^{-2k\vert y \vert}\left(-dt^2 + d\vec{x}^2  \right) + dy^2,
\label{Metric1}
\end{equation}
where $k$ is related to the anti-de Sitter curvature radius $r_{AdS}$ by $k=1/r_{AdS}$, and
$e^{-2k\vert y \vert}$ is the so-called warp factor. The five-dimensional
bulk $(\mathcal{M},g_{MN})$ is asymptotically anti-de Sitter, and the four-dimensional 
brane $(\Sigma,h_{\mu\nu})$ is located at $y=0$ in the extra dimension. As we will see, the brane spacetime
could be asymptotically flat or not.

\begin{figure}
\begin{centering}
\includegraphics[trim=1cm 0cm 1cm 0cm, clip=true,scale=0.6]{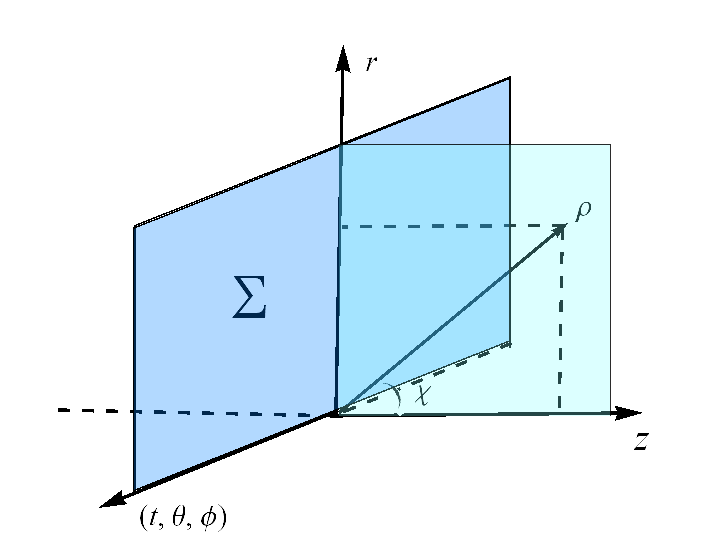}
\par\end{centering}
\caption{Representation of the four-dimensional brane spacetime, indicated 
by $\Sigma$, embedded in the five-dimensional bulk 
with suppression of one spacial dimension. The brane is located at $z=0$. 
The coordinates $(t,r,\theta,\phi)$ are the brane coordinates.} 
\label{Diagram}
\end{figure}

The metric (\ref{Metric1}), in the flat coordinates, is clearly conformally flat, that is, making the use of the
new coordinate $z=sgn (y)(e^{k\vert y\vert}-1)/k$, one has  
\begin{equation}
ds^2=\frac{1}{\left(1+k\vert z \vert \right)^2}\left(-dt^2+dr^2+r^2d\Omega_2^2+dz^2 \right),
\label{Metric2}
\end{equation}
where $d\Omega_2^2=d\theta^2+\sin^2 \theta d\phi^2$ is the line-element of a unit two-sphere. 
Notice that the brane is still located at $y=0$ or  $z=0$.
The Nakas-Kanti  approach \cite{Nakas} pursues the five-dimensional spacetime that generates known metrics
on the brane. In order to build the bulk spacetime, one should impose the spherical symmetry (in our case)
 in the bulk. Accordingly, the following change of coordinates is made:
\begin{equation}
r= \rho \sin \chi  \hspace{0.5cm} \mbox{and} \hspace{0.5cm} z= \rho \cos \chi ,
\label{r,z}
\end{equation}
with $\chi \in [0,\pi]$. From these coordinate transformations, the bulk metric (\ref{Metric2}) is written as
\begin{equation}
ds^2=\frac{1}{\left(1+k\rho \vert \cos\chi \vert \right)^2}\left(-dt^2+d\rho^2+\rho^2d \Omega_3^2 \right),
\label{Metric3}
\end{equation}
where $d\Omega_3^2=d\chi^2+\sin^2 \chi d\theta^2+\sin^2 \chi \sin^2 \theta d\phi^2$ is the line-element from a unit three-sphere. Also, the inverse transformations of (\ref{r,z}) are written as
\begin{equation}
\rho(r,z) =\left(r^2+z^2\right)^{\frac{1}{2}} \hspace{0.5cm} \mbox{and} \hspace{0.5cm} \tan \chi =\frac{r}{z}.
\label{rho}
\end{equation}
It worth noting  that the new radial coordinate $\rho$ brings the bulk extra coordinate $z$ 
and the brane radial coordinate $r$, it ranges
from 0 to $\infty$. The angular coordinate $\chi$ indicates both sides of the brane. 
For $[0,\pi/2)$, $\chi$ gives us the \enquote{right} side (positive values of $z$) of the brane, 
and for $(\pi/2,\pi]$, we have the \enquote{left} side (see Fig. \ref{Diagram}). 
Most importantly, due to the Randall-Sundrum II $\mathbb{Z}_2$ 
symmetry in the bulk,  the points $z$ and $-z$ are very equivalent.  
Therefore, for the sake of simplicity, the modulus of $y$ in Eq. (\ref{Metric3}) could be dropped in some calculations.

In order to generate five-dimensional black holes or wormholes, we should replace the metric 
elements $-dt^2 + d\rho^2$ in Eq. (\ref{Metric3}) by any known spacetime geometry. 
As in this article the subject is wormhole solutions, I use the general spherically symmetric $\textit{Ansatz}$
in (\ref{Metric3}), that is,
\begin{equation}
ds^2=\frac{1}{\left(1+k\rho  \cos\chi  \right)^2}\left(-A(\rho) dt^2+\frac{d\rho^2}{B(\rho)} +\rho^2d \Omega_3^2 \right).
\label{Metric4}
\end{equation}
Adopting the two metric functions $A(\rho)$ and $B(\rho)$ of known four-dimensional
 spacetime metrics yields five-dimensional black holes \cite{Nakas,Nakas:2021srr,Neves:2021dqx}
  or, as we will see, wormholes. 
 In this study, $A(\rho)$ and $B(\rho)$ will be the same of the Morris-Thorne \cite{MT} 
 and Molina-Neves \cite{MN} wormholes.

A last change of coordinates is useful to characterize horizons on the brane and, specially, in the bulk.
The metric (\ref{Metric4}), in the original $(t,r,\theta,\phi,y)$ coordinates, is written as 
\begin{align}
 ds^2 &= e^{-2k\vert y \vert}\bigg[ -A(r,y)dt^2+\bigg(\frac{r^2}{B(r,y)}+z(y)^2 \bigg) \nonumber \\
& \times \frac{dr^2}{r^2+z(y)^2}+  \bigg(\frac{1}{B(r,y)}-1 \bigg)\frac{2rz(y)e^{k\vert y \vert}}{r^2+z(y)^2}drdy \nonumber \\ 
 & +  r^2d\Omega_2^2 \bigg] + \bigg(r^2+\frac{z(y)^2}{B(r,y)}\bigg)  \frac{dy^2}{r^2+z(y)^2}.
\label{Metric5}
\end{align}
This metric shows easily whether or not any type of horizon extends to the extra dimension $y$. 
Because zeros of $g^{rr}=B(r,y)=0$ give us event horizons and the wormhole throat localization. 
In order to be in agreement with the anti-de Sitter spacetime, off-diagonal
metric terms will be zero as $y\rightarrow \infty$.

In the Nakas-Kanti approach, the bulk geometry is solution of the five-dimensional gravitational
field equations
\begin{equation}
G_{MN}=\kappa_5^2 T_{MN}^{(B)},
\label{5d-FE}
\end{equation}
where the parameter $\kappa_5^2=8\pi G_5$ is the gravitational constant in five dimensions, and the tensors 
$G_{MN}$ and $T_{MN}^{(B)}$ stand for the five-dimensional Einstein tensor  and the bulk 
energy-momentum tensor, respectively.  It is worth noticing that the field equations
do not present the five-dimensional cosmological constant. Indeed, the tensor $T_{MN}^{(B)}$
represents an exotic fluid that simulates the cosmological constant, providing then 
the anti-de Sitter asymptotic behavior in the bulk and, as will see,  surprisingly, inducing even a positive cosmological
constant on the four-dimensional brane.

\section{The brane equations}
\label{Sec-III}
On the brane, the field equations are neither the five-dimensional equations (\ref{5d-FE}) nor the Einstein equations.
By using the Gauss and Codacci equations, Shiromizu, Maeda and Sasaki \cite{Shiromizu} 
calculated the effective field equations on the brane. Such modified field equations possess new terms that 
measure the bulk influence on the brane. 
As previously mentioned, the four-dimensional brane $(\Sigma, h_{MN})$ is
embedded into a five-dimensional bulk $(\mathcal{M},g_{MN})$, consequently a normal and unit vector to the brane
is written as $n^M= \delta^M_{\ y}$. Therefore, the induced metric on the brane reads $h_{MN}=g_{MN}-n_{M}n_{N}$. Following Nakas and Kanti \cite{Nakas}, 
the total energy-momentum tensor could be written as
\begin{equation}
T_{MN}=T^{(B)}_{MN}+ \delta^{\mu}_{M} \delta^{\nu}_{N}T^{(br)}_{\mu\nu} \delta (y),
\label{T_total}
\end{equation}
where the delta  $\delta (y)$ asks for a junction condition, and 
the brane energy-momentum tensor, following many brane-world scenarios \cite{Shiromizu}, is given by
\begin{equation}
T^{(br)}_{\mu\nu}=-\lambda h_{\mu\nu}+\tau_{\mu\nu}.
\label{T_brane}
\end{equation}
The constant $\lambda$ is the brane tension, which is some sort of  vacuum energy on the brane, and the tensor 
$\tau_{\mu\nu}$ represents all matter fields or sources exclusively on the brane. For the two wormholes studied here,  
 $\tau_{\mu\nu}=0$, and induced metrics on the brane will be caused by the bulk 
influence on the brane. It is important to note that the total energy-momentum tensor (\ref{T_total}) is
different from the one studied in Ref. \cite{Shiromizu}. The bulk energy-momentum tensor adopted here
 is not described by a cosmological constant term. As I said, the bulk asymptotic anti-de Sitter behavior (for
 large values of the extra coordinate) and even the de Sitter brane in the Molina-Neves wormhole 
 are supported by the exotic fluid in the bulk.

In order to show that the brane adopted in this article has no fields or particles that live just on the four-dimensional 
spacetime or, equivalently, $\tau_{\mu\nu}=0$, 
the Israel junction condition \cite{Israel:1966rt} at $y=0$ is required. Consequently, the following 
results for the extrinsic curvature $K_{\mu\nu}$ and the induced metric $h_{\mu\nu}$ are used:   
\begin{align}
\left[K_{\mu\nu} \right] & =  -\kappa_5^2 \left(T^{(br)}_{\mu\nu} -\frac{1}{3}h_{\mu\nu}T^{(br)} \right), 
\label{K1} \\
[h_{\mu\nu}] & = 0,
\end{align}
where the bracket notation means
\begin{equation}
\left[X \right]= \lim_{y\rightarrow 0^+}X- \lim_{y \rightarrow 0^-}X=X^+-X^-,
\end{equation}
and $T^{(br)}$ is the trace of the four-dimensional energy-momentum
tensor on the brane. As the extrinsic curvature is defined as
\begin{equation}
K_{\mu\nu}= h^A_{\ \mu} h^B_{\ \nu} \nabla_A n_B,
\label{K2}
\end{equation}
in which $n^{A}=(0,0,0,0,1/ \sqrt{g_{yy}(y=0)})$, and 
the induced metric $h_{\mu\nu}$ on the brane in the $(t,r,\theta,\phi)$ coordinates is written as
\begin{equation}
ds^2= h_{tt} dt^2+ h_{rr}dr^2 + h_{\theta\theta}d\theta^2 + h_{\phi\phi}d\phi^2,
\label{Induced_metric}
\end{equation}
(which satisfies $[h_{\mu\nu}]=0$), the metric (\ref{Metric5}) will result in
\begin{equation}
K_{\mu\nu}=-k\frac{d\vert y \vert}{dy}h_{\mu\nu},
\label{K}
\end{equation}  
since
\begin{equation}
\frac{\partial A(r,y)}{\partial y} \bigg \vert _{y \rightarrow 0} = \frac{\partial B(r,y)}{\partial y} \bigg \vert _{y \rightarrow 0}= 0,
\end{equation}
which is the case for the wormhole geometries studied here. Therefore, it follows
that $K=-4k \frac{d\vert y \vert}{dy}$. With the extrinsic curvature calculated, the definition (\ref{K1})
yields
\begin{equation}
T^{(br)}=\frac{3}{\kappa_5^2}[K].
\end{equation}
Thus, we can rewrite Eq (\ref{K1}) as
\begin{equation}
T^{(br)}_{\mu\nu}=-\frac{1}{\kappa_5^2}\left(\left[K_{\mu\nu} \right]-\left[ K\right]h_{\mu\nu} \right)=-\frac{6k}{\kappa_5^2}h_{\mu\nu},
\label{Trace}
\end{equation} 
where the $\mathbb{Z}_2$ symmetry  was applied. Consequently, from Eq. (\ref{Trace}) and Eq. (\ref{T_brane}),
one shows that $\tau_{\mu\nu}=0$ and that \textit{there are no sources or matter fields whose origin is on the brane}. 
Moreover, as a consequence, the brane tension is related to $k$ (and the anti-de Sitter curvature) 
by means of $\lambda=6k/\kappa_5^2>0$.
Then the induced geometries on the brane, whether black holes \cite{Nakas,Nakas:2021srr,Neves:2021dqx} 
or wormholes as this article studies,
are generated only by the bulk influence on the brane. 

The effective field equations on the brane, deduced by Shiromizu, Maeda and Sasaki \cite{Shiromizu}, 
quantifies the bulk influence on the brane. Such field equations are given by
\begin{align}
G_{\mu\nu}=& \frac{2\kappa_5^2}{3}\left[h^{M}_{\ \mu} h^{N}_{\ \nu} T^{(B)}_{MN}+\left( T^{(B)}_{MN}n^M n^N -\frac{T^{(B)}}{4}  \right) h_{\mu\nu} \right] \nonumber \\
&  + KK_{\mu\nu}- K^{\ \alpha}_{\mu}K_{\nu \alpha}-\frac{1}{2}h_{\mu\nu}\left(K^2-K^{\alpha \beta}K_{\alpha \beta} \right) \nonumber \\
& - E_{\mu\nu},
\label{Field_equations}
\end{align}
where $G_{\mu\nu}$ is the four-dimensional Einstein tensor, $T^{(B)}$ is the trace of the 
bulk energy-momentum tensor, and $E_{\mu\nu}$ is the so-called \enquote{electric} part
of the five-dimensional Weyl tensor $(C^A_{\ \ BCD})$ projected on the brane. It is explicitly written as
\begin{equation}
E_{\mu\nu}= C^A_{\ \ BCD}n_An^C h^{B}_{\ \mu} h^{D}_{\ \nu}.
\label{E_definition}
\end{equation}
An important result from this tensor is $E^{\mu}_{\ \mu}=0$, that is, it is a traceless tensor. 

By using Eq. (\ref{K}), the terms of the effective field equations (\ref{Field_equations}) that depend on the extrinsic 
curvature are
\begin{align}
& KK_{\mu\nu}- K^{\ \alpha}_{\mu}K_{\nu \alpha}-\frac{1}{2}h_{\mu\nu}\left(K^2-K^{\alpha \beta}K_{\alpha \beta} \right) = 8\pi G \tau_{\mu\nu}  \nonumber \\
&+\kappa_5^4 \left( \pi_{\mu\nu}-\frac{\lambda^2}{12} h_{\mu\nu} \right),
\label{KK}
\end{align} 
where $G=\kappa_5^4 \lambda/48\pi=1$ stands for the effective 
gravitational constant on the brane, and
\begin{equation}
\pi_{\mu\nu}=-\frac{1}{4}\tau_{\mu\alpha} \tau^{\ \alpha}_{\nu}+\frac{1}{12}\tau \tau_{\mu\nu}+\frac{1}{8}\tau_{\alpha \beta}\tau^{\alpha \beta} h_{\mu\nu} -\frac{\tau^2}{24} h_{\mu\nu}.
\label{Tau}
\end{equation}
Inserting the results (\ref{KK})-(\ref{Tau}) into the field equations (\ref{Field_equations}), 
then we have the final form of the effective equations that depend on a new tensor 
$T_{\mu\nu}^{(\text{eff})}$, namely 
\begin{equation}
G_{\mu\nu}= \kappa_5^2 k T_{\mu\nu}^{(\text{eff})} - 3k^2 h_{\mu\nu}  - E_{\mu\nu},
\label{Field_eq2}
\end{equation}
where $\tau_{\mu\nu}=0$, consequently $\pi_{\mu\nu}=0$, and $\lambda=6k/\kappa_5^2>0$. 
The new energy-momentum tensor (diagonal tensor) $T_{\mu\nu}^{(\text{eff})}$ is some sort of an
effective energy-momentum tensor on the brane. Together with $E_{\mu\nu}$, the tensor 
$T_{\mu\nu}^{(\text{eff})}$ measures the influence of the bulk fields on the brane. It is defined as
\begin{equation}
T_{\mu\nu}^{(\text{eff})} = \frac{2}{3k}\left[T^{(B)}_{\mu\nu}+\left( T^{(B)}_{yy} -\frac{T^{(B)}}{4}  \right) h_{\mu\nu} \right] \Bigg|_{y \rightarrow 0},
\label{T_eff}
\end{equation}
calculated at $y=0$. Therefore, any spacetime metric induced on the brane should \textit{necessarily} 
be solution of (\ref{Field_eq2}). 
And, as we will see, the two wormhole solutions studied here satisfy this necessary condition.

\section{Wormholes in the brane context}
\label{Sec-IV}

\subsection{Morris-Thorne wormhole}
\begin{figure}
\begin{centering}
\includegraphics[trim=3.2cm 0cm 2.2cm 0cm, clip=true,scale=0.54]{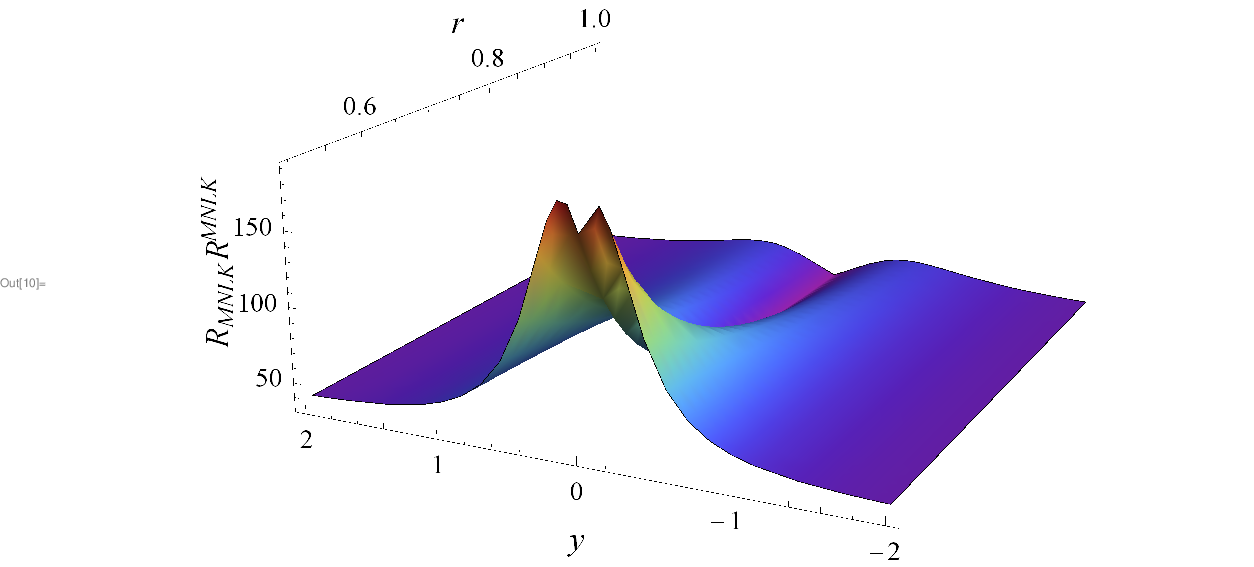}
\par\end{centering}
\caption{The Kretschmann scalar $R_{MNLK}R^{MNLK}$ for the
five-dimensional Morris-Thorne wormhole indicates that spacetime is regular in the extra dimension $y$.
In this graphic, one uses $k=1$ and $b_0=0.5$.} 
\label{Scalar-MT}
\end{figure}

The class of four-dimensional Morris-Thorne wormholes in the ($t,r,\theta,\phi$) coordinates is given by
\begin{equation}
ds^2= -e^{2\Phi(r)}dt^2+\left(1-\frac{b(r)}{r}\right)^{-1}dr^2+r^2d \Omega_{2}^{2}.
\label{MT}
\end{equation}
Here I will focus on the simplest case, a case in which spacetime is horizonless and without tidal forces. 
Thus, the metric functions are
\begin{align}
\Phi (r) & =  0, \label{MT-I} \\
b(r)  & =  \sqrt{b_0 r}, \label{MT-II}
\end{align} 
where $r_{\text{thr}}=b_0$ is the wormhole  throat coordinate in four dimensions. 
Following the Nakas-Kanti approach, then one inserts the metric (\ref{MT}) with above functions into (\ref{Metric4}), 
where in this case
\begin{equation}
A(\rho) = 1 \hspace{0.5cm} \mbox{and} \hspace{0.5cm} B(\rho) = 1-\sqrt{\frac{b_0}{\rho}}.
\label{B_rho}
\end{equation}
 Thus, one calculates some scalars in order to check the asymptotic behavior of the 
five-dimensional spacetime. The Ricci scalar or scalar curvature, for example, is
\begin{equation}
\lim_{\rho \rightarrow \infty}R=-20k^2=-\frac{20}{r_{AdS}^2},
\end{equation}
and shows us that the five-dimensional geometry (\ref{Metric4}) 
with the elements or functions of the Morris-Thorne  wormhole 
is asymptotically anti-de Sitter in the brane context adopted in this article.  
Other scalars, like $R_{MN}R^{MN}$ and $R_{MNLK}R^{MNLK}$ (called Kretschmann scalar),
 indicate the same asymptotic behavior, i.e., 
\begin{align}
& \lim_{\rho \rightarrow \infty}R_{MN}R^{MN} = 80k^4, \\
& \lim_{\rho \rightarrow \infty}R_{MNLK}R^{MNLK} = 40k^4.
\end{align}
Another interesting limits are given at the throat on the brane, $ \rho_{\text{thr}}=(b_0,0)$, namely
\begin{align}
& \lim_{\rho \rightarrow \rho_{\text{thr}}} R = -20k^2 +\frac{9}{2b_0^2}, \\
& \lim_{\rho \rightarrow \rho_{\text{thr}}} R_{MN}R^{MN} =  80k^4 -\frac{36k^2}{b_0^2}+\frac{39}{4b_0^4}, \\
& \lim_{\rho \rightarrow \rho_{\text{thr}}} R_{MNLK}R^{MNLK} = 40k^4  -\frac{18k^2}{b_0^2}+\frac{51}{4b_0^4}.
\end{align} 
As we can see in Fig. \ref{Scalar-MT}, the five-dimensional metric is regular everywhere. 

In order to see the horizons existence (or not) in the extra dimension, we should adopt the metric (\ref{Metric5}) 
and the coordinates $r$ and $y$ instead of $\rho$. The metric (\ref{Metric5}) give us the following elements in the
Morris-Thorne case:
\begin{align}
A(r,y) & = 1, \label{A-MT} \\
B(r,y) & = 1- \frac{\sqrt{k  b_0}}{[\left(e^{k \vert y \vert}-1\right)^2 + k^2r^2]^{\frac{1}{4}}}.
\label{B-MT}
\end{align}
As $g^{rr}=B(r,y)=0$ provides the horizon localization, the metric function (\ref{B-MT}) 
has just one zero or root,
which is the wormhole throat at $(r_{\text{thr}},y_{\text{thr}})$. 
Then the five-dimensional Morris-Thorne wormhole is traversable.
Like the event horizon of the  Schwarzschild black hole  \cite{Nakas} and regular black holes \cite{Neves:2021dqx},
 the wormhole throat extends to the
extra dimension $y$. From (\ref{B-MT}), the wormhole throat constraint is
\begin{equation}
r_{\text{thr}}^2 =b_0^2 -\frac{\left(e^{k \vert y_{\text{thr}} \vert}-1\right)^2}{k^2}.
\label{Constraint}
\end{equation} 
Interestingly, assuming that the coordinate $\rho \in [b_0,+\infty)$ and $r \in [0,+\infty)$,
 the wormhole throat constraint is just $r_{\text{thr}}^2+z_{\text{thr}}^2=b_0^2$, which is 
 equivalently to (\ref{Constraint}). Also, with these assumptions, the definitions (\ref{r,z}) are valid 
 because $\chi \in [0,\pi]$ is still true, and the limit $r_{\text{thr}}=b_0$ is satisfied on the brane.\footnote{Recently, Pappas and Nakas \cite{Pappas:2024qwm} explore the details of 
 of these assumptions regarding the radial coordinates $\rho$ and $r$  in order to build consistent embedding of wormholes into the bulk.}
 
It is worth noticing that with $\rho \in [b_0,+\infty)$ the metric signature is correct in the bulk.
 From (\ref{B_rho}), we see that $B(\rho)\geq 0$ through the five-dimensional spacetime.

The matter content in the bulk is investigated with the five-dimensional gravitational field equations (\ref{5d-FE}).
From the $(t,\rho,\theta,\phi,\chi)$ coordinates, the bulk fluid could be described by the energy-momentum
tensor components
\begin{align}
T^{(B)t}_{\ \ \ \ \  t}  = & \ \frac{1}{\kappa_{5}^{2}} \left[6 k^2 + \frac{\sqrt{b_0 \rho }}{4} \left(\frac{15 k \cos \chi}{\rho ^2}-\frac{9}{\rho ^3}\right)\right], \label{T1-MT} \\
T^{(B)\rho}_{\ \ \ \ \  \rho}  = & \  \frac{1}{\kappa_{5}^{2}} \left[6k^2 + 3\sqrt{b_0 \rho } \left(\frac{k \cos \chi}{\rho ^2}-\frac{1}{\rho ^3}\right)\right],  \label{T2-MT}  \\
T^{(B)\theta}_{\ \ \ \ \ \theta}  = & \  T^{(B)\phi}_{\ \ \ \  \ \phi} = T^{(B)\chi}_{\ \ \ \ \  \chi} = \frac{1}{\kappa_{5}^{2}} \Bigg[  6k^2 \nonumber \\
&  - \frac{\sqrt{b_0 \rho}}{4} \left( \frac{5 k^2 \cos ^2 \chi}{ \rho }-\frac{17 k \cos \chi}{ \rho ^2} + \frac{2}{\rho ^3}\right) \Bigg]. \label{T3-MT} 
\end{align}
As we can see, using $(t,\rho,\theta,\phi,\chi)$ coordinates, the bulk energy-momentum tensor is diagonal, that is, 
 $T^{(B)\mu}_{\ \ \ \ \ \nu}=diag(-\rho_E, p_1, p_2,p_2,p_2)$, where $\rho_E$ 
 is the energy density, and $p_1$ and $p_2$ are pressures of the fluid, which in this 
 case is an anisotropic fluid due to $p_1\neq p_2$. As expected, the limit of the energy-momentum tensor components
 illustrates the asymptotic anti-de Sitter behavior of the five-dimensional spacetime, i.e., 
\begin{align}
\lim_{\rho \rightarrow \infty} \rho_E= & -\frac{6k^2}{\kappa_5^2}= \frac{\Lambda_{5d}}{\kappa_5^2}, \label{Limit1} \\
\lim_{\rho \rightarrow \infty} p_1= & \lim_{\rho \rightarrow \infty} p_2= \frac{6k^2}{\kappa_5^2}= - \frac{\Lambda_{5d}}{\kappa_5^2},
\label{Limit2}
\end{align}
in which $\Lambda_{5d}$ is negative and is the \enquote{effective} five-dimensional cosmological constant. As the field
equations (\ref{5d-FE}) for the bulk  do not possess the cosmological constant term, the fluid 
given by Eqs. (\ref{T1-MT})-(\ref{T3-MT}) simulates the five-dimensional anti-de Sitter spacetime at this limit.

\begin{figure}
\begin{centering}
\includegraphics[trim=0.3cm 0.3cm 1.4cm 0cm, clip=true,scale=0.55]{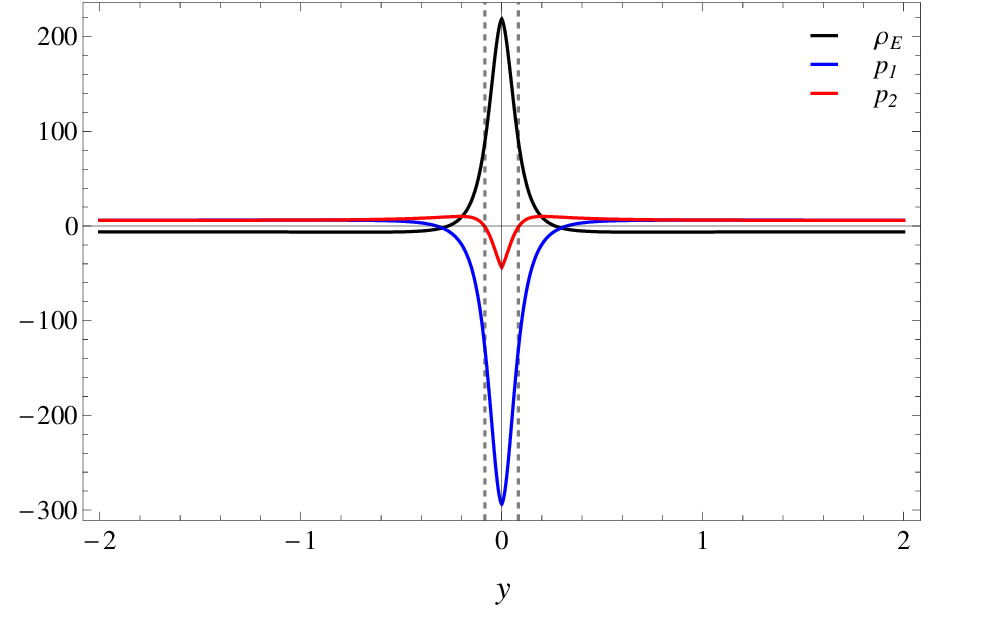}
\par\end{centering}
\caption{Components of the bulk energy-momentum tensor and the weak energy violation close to the brane (located
 at $y=0$)  for the Morris-Thorne case. In this graphic, one adopts $\kappa_5=k=1$, $b_0=0.1$, $r=b_0$, and 
 $r_{\text{thr}}=0.05$. The vertical dashed lines ($ y_{\text{thr}} = \pm 0.08$) indicate the wormhole throat in the extra dimension.}
\label{Energy-MT}
\end{figure}

On the other hand, the dependence of the energy-momentum tensor components on the extra dimension 
 is better visualized in the $(t,r,\theta,\phi,y)$ coordinates. In these coordinates, the energy density and pressures are
\begin{align}
\rho_E &  = -\frac{6k^2}{\kappa_5^2}\Bigg \{1- \frac{\sqrt{b_0 r}\left(1-\frac{5}{8}e^{k\vert y \vert} \right)}{[(e^{k \vert y \vert} -1)^2 + k^2 r^2]^{\frac{5}{4}}} \Bigg \}, \\
p_1 & = \frac{6k^2}{\kappa_5^2}\Bigg \{1- \frac{\sqrt{b_0 r}\left(1-\frac{1}{2}e^{k\vert y \vert} \right)}{[(e^{k \vert y \vert} -1 )^2 + k^2 r^2]^{\frac{5}{4}}} \Bigg \}, \\
p_2 &  = \frac{6k^2}{\kappa_5^2}\Bigg \{1- \frac{\sqrt{b_0 r}\left(1-\frac{27}{24}e^{k\vert y \vert} +\frac{5}{24}e^{2k\vert y \vert} \right)}{[(e^{k \vert y \vert} -1)^2 + k^2 r^2]^{\frac{5}{4}}} \Bigg \}. \\
\end{align}
In these coordinates, the energy-momentum tensor is not diagonal anymore. There are two off-diagonal
components $T^{(B)r}_{\ \ \ \ \  y}$ and $T^{(B)y}_{\ \ \ \ \ r}$, which are zero for the limit $y \rightarrow \infty$ 
and, or course, for $y=0$. 
As we can see in Fig. \ref{Energy-MT}, energy conditions could be violated.
In particular, close to the brane (or close to the wormhole throat), for small values of $y$, 
the weak energy condition is violated, namely 
\begin{equation}
\rho_E +p_1 \simeq - \frac{3}{4\kappa_5^2}\sqrt{\frac{b_0}{r^5}}\left(1+k \vert y \vert \right) < 0.
\end{equation}

With the five-dimensional wormhole metric calculated, given by the metric functions (\ref{A-MT}) and (\ref{B-MT}), 
it is time to verify whether or not the Morris-Thorne
wormhole could be a wormhole solution on the brane. First of all, let us assume that 
the induced metric on the brane, namely $h_{\mu\nu}$, is
the Morris-Thorne wormhole (\ref{MT}). Thus, the \enquote{electric} part of the 5-dimensional Weyl tensor is
\begin{equation}
E_{\mu\nu} \bigg \vert_{y \rightarrow 0 }= \frac{5}{24}\sqrt{\frac{b_0}{r^5}} \left(\begin{array}{cccc}
-h_{tt}\\
 &-h_{rr} \\
 &  & h_{\theta \theta} \\
 &  &  & h_{\phi \phi}
\end{array}\right),
\label{E-MT}
\end{equation}
where the condition $E^{\mu}_{\ \mu}=0$ clearly is satisfied. Now, the tensor (\ref{T_eff}) 
should be evaluated. With the assumed bulk and induced metrics, we arrive at the following tensor:
\begin{align}
T_{\mu\nu}^{(\text{eff})} = & \ \frac{1}{\kappa_5^2 k}\Bigg[\sqrt{\frac{b_0}{r^5}} \left(\begin{array}{cccc}
-\frac{17}{24} h_{tt}\\
 &- \frac{29}{24} h_{rr} \\
 &  & \frac{11}{24} h_{\theta \theta} \\
 &  &  & \frac{11}{24} h_{\phi \phi}
\end{array}\right) \nonumber \\
& + 3k^2 h_{\mu\nu} \Bigg].
\label{T-MT}
\end{align}
Now, inserting Eqs. (\ref{E-MT}) and (\ref{T-MT}) into the field equations on the brane, given by 
Eq. (\ref{Field_eq2}), we have the following
expression for the four-dimensional Einstein tensor:
\begin{equation}
G_{\mu\nu} =  \sqrt{\frac{b_0}{r^5}} \left(\begin{array}{cccc}
 -\frac{1}{2} h_{tt}\\
 & - h_{rr}  \\
 &  &  \frac{1}{4} h_{\theta \theta}  \\
 &  &  &  \frac{1}{4} h_{\phi \phi} 
\end{array}\right),
\label{G-MT}
\end{equation}
which is the Einstein tensor of the Morris-Thorne geometry (\ref{MT}) in the general relativity context. 
Therefore, the Morris-Thorne wormhole could be induced on the brane and could be solution of the
four-dimensional effective field equations on the brane. However, contrary to the general relativity  context
with the Einstein field equations, the effective field equations tell us that the Morris-Thorne wormhole 
is supported by an exotic fluid whose origin is in the five-dimensional spacetime.

\subsection{Molina-Neves wormhole}
The Molina-Neves wormhole \cite{MN} was built in a brane-world scenario. 
It is a four-dimensional wormhole in an 
asymptotically de Sitter brane. As mentioned in Introduction, this wormhole metric was obtained without knowing the 
five-dimensional bulk configuration. In this sense, the goal of this subsection is finding the bulk spacetime
in which the Molina-Neves solution is the induced metric on the four-dimensional brane.

The Molina-Neves spacetime in the $(t,r,\theta,\phi)$  coordinates is given by
\begin{equation}
ds^2= - e^{2\Phi (r)}dt^2 +\left[e^{2\Phi (r)}\left(1- \frac{b(r)}{r} \right) \right]^{-1}dr^2 +r^2 d\Omega_{2}^{2},
\label{MN}
\end{equation}
where in this case
\begin{align}
\Phi (r) = & \frac{1}{2} \ln  \left(1-\frac{r^2}{r_c^2} \right), \label{phi}  \\
b(r)= &  \frac{C}{\left( r^2-r_0^2\right)^{\frac{3}{2}}}.\label{b}
\end{align} 
The parameters $r_0=\sqrt{2/ \Lambda_{4d}}$ and $r_c=\sqrt{3/ \Lambda_{4d}}$ depend on the
four-dimensional cosmological constant $\Lambda_{4d}$, which also will be
conceived of as a positive and \enquote{effective} cosmological constant.
In order to generate wormholes, the constant $C$ should be positive. It has 
dimension $L^4$ in geometrized units.  Zeros of $h^{rr}=0$ give us 
the cosmological horizon on the brane, $r= r_c$, and the wormhole throat $r=b_0$. It is worth emphasizing
that the radial coordinate is defined in the interval
\begin{equation}
r_0<b_0 \leq r \leq r_c,
\end{equation}
in which the spacetime metric is regular for $r \neq r_0$.

\begin{figure}
\begin{centering}
\includegraphics[trim=3.5cm 0cm 2.8cm 0cm, clip=true,scale=0.6]{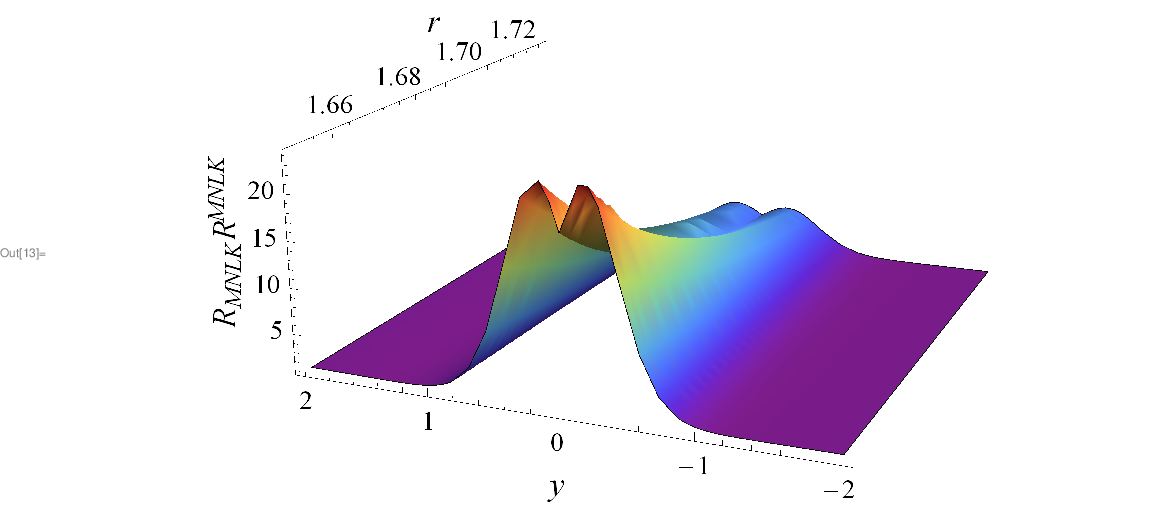}
\par\end{centering}
\caption{The Kretschmann scalar $R_{MNLK}R^{MNLK}$ for the
five-dimensional Molina-Neves wormhole indicates that spacetime is also regular in the extra dimension $y$.
In this graphic, one uses $C=\Lambda_{4d}=1$, $\Lambda_{5d}=-1$, $k=1/\sqrt{2}$, and $b_0=1.65$.}
\label{Scalar-MN}
\end{figure}

\begin{figure}
\begin{centering}
\includegraphics[trim=0.1cm 0.2cm 0.3cm 0cm, clip=true,scale=0.6]{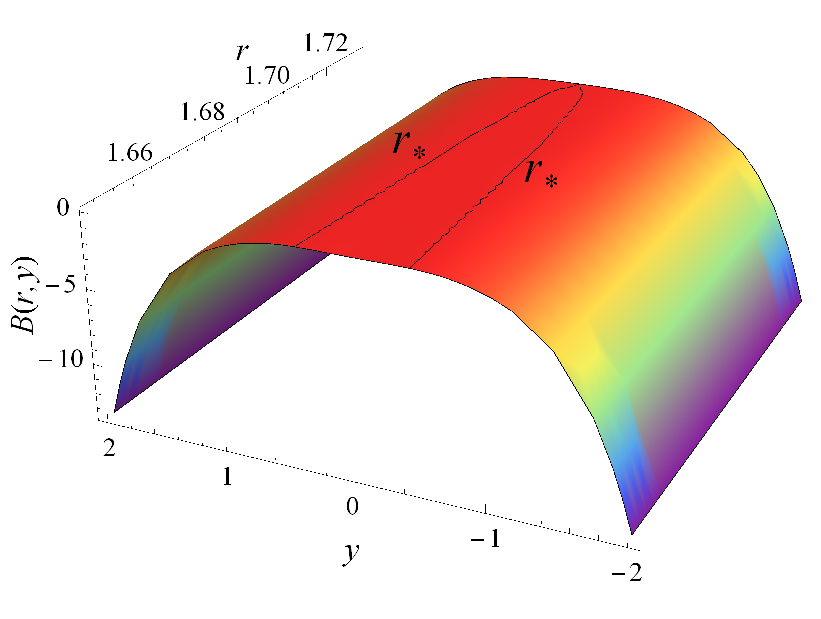}
\par\end{centering}
\caption{$B(r,y)$  for the Molina-Neves metric. The black line over the surface indicates 
that the cosmological horizon $r_* = (r_c,y_c)$ extends to the extra dimension $y$. In this graphic, one adopts
$\Lambda_{4d} =1$ and $\Lambda_{5d}=-4$, consequently $k=C=1$.} 
\label{B}
\end{figure}

Like the Morris-Thorne case, in order to get a five-dimensional wormhole one should insert (\ref{MN}) 
into Eq. (\ref{Metric4}). In this case
\begin{equation}
A(\rho)= e^{2\Phi (\rho)} \hspace{0.5cm} \mbox{and} \hspace{0.5cm} B(\rho)= e^{2 \Phi (\rho)}\left(1- \frac{b(\rho)}{\rho} \right),
\label{B_rho2}
\end{equation}
where the functions $\Phi (\rho)$ and $b(\rho)$ are given by Eqs. (\ref{phi})-(\ref{b}) 
replacing the coordinate $r$ by $\rho$.
Once again, in order to see the anti-de Sitter behavior of the bulk, scalars should be calculated. For this case, the 
scalar curvature reads
\begin{equation}
\lim_{\rho \rightarrow \infty}R = -20k^2+\frac{20}{r_c^2}.
\end{equation}
Therefore, in agreement with Ref. \cite{Nakas:2023yhj}, 
where the de Sitter geometry was briefly commented in the Nakas-Kanti 
approach, one assumes that
\begin{equation}
k^2= \frac{\Lambda_{4d}}{3} - \frac{\Lambda_{5d}}{6}.
\end{equation}
Thus, the curvature parameter in bulk metric (\ref{Metric1}) also depends on the 
\enquote{effective} four-dimensional cosmological constant.
With this definition of $k$, all three scalars are in agreement with an asymptotically 
  anti-de Sitter space in five dimensions, namely
\begin{align}
& \lim_{\rho \rightarrow \infty}R  = \frac{10}{3} \Lambda_{5d}, \\
& \lim_{\rho \rightarrow \infty}R_{MN}R^{MN}  = \frac{20}{9} \Lambda_{5d}^2, \\
& \lim_{\rho \rightarrow \infty}R_{MNLK}R^{MNLK} =  \frac{10}{9} \Lambda_{5d}^2.
\end{align}   
These limits taken at the wormhole throat on the brane, $\rho_{\text{thr}}= (b_0,0)$, are
\begin{align}
& \lim_{\rho \rightarrow \rho_{\text{thr}}} R  \sim \frac{1}{\left(b_0^2 - r_0^2 \right)^{\frac{5}{2}}}, \\
& \lim_{\rho \rightarrow \rho_{\text{thr}}} R_{MN}R^{MN}   \sim \frac{1}{\left(b_0^2 - r_0^2 \right)^5}, \\
& \lim_{\rho \rightarrow \rho_{\text{thr}}} R_{MNLK}R^{MNLK}  \sim \frac{1}{\left(b_0^2 - r_0^2 \right)^5}.
\end{align} 
As we can see, once again, the five-dimensional metric is regular at the throat. 
Also according to Fig. \ref{Scalar-MN}, the bulk metric is regular everywhere including the wormhole
throat. By making use of the $(t,r,\theta,\phi,y)$ 
coordinates, the cosmological horizon and the throat localization will be more clear. Thus, with
\begin{align}
A(r,y) & = 1-\frac{r^2+z(y)^2}{r_{c}^2}, \label{A-MN}\\
B(r,y) & = A(r,y)\Bigg(1- \frac{C}{\left[r^2+z(y)^2 \right]^{\frac{1}{2}} \left[r^2+z(y)^2-r_0^2 \right]^{\frac{3}{2}}} \Bigg),\label{B-MN}
\end{align}
in which $z(y)=sgn (y)(e^{k\vert y\vert}-1)/k$, the zeros of $g^{rr}=B(r,y)=0$ provide the localization of 
such important surfaces. The cosmological horizon $r_* = (r_c,y_c)$ constraint is written as
\begin{equation}
r_c^2 = \frac{3}{\Lambda_{4d}} -\frac{\left(e^{k \vert y_c \vert} -1 \right)^2}{k^2}.
\end{equation}
Then the cosmological horizon $r_* = (r_c,y_c)$ extends to the 
bulk and decays exponentially in the extra coordinate 
like the event horizon of black holes in this very same context  \cite{Nakas,Nakas:2021srr,Neves:2021dqx} (see Fig. \ref{B}).
 As we can see, the value of the radial coordinate for the cosmological horizon 
 on the brane is restored for $y=0$. In particular,
 the maximum value of the coordinate $y_c$ is given by
 \begin{equation}
 \vert y_c \vert = \frac{1}{k}\ln \left(1+ k \sqrt{\frac{3}{\Lambda_{4d}}-b_0^2} \right).
 \label{y_c}
\end{equation}   
According to Nakas and Kanti \cite{Nakas}, this very feature gives to the horizons a \enquote{pancake} shape in the extra
dimension. 

Once again,  assuming that the coordinate $\rho \in [b_0,+\infty)$ and $r \in [0,+\infty)$,
the positivity of the metric terms (\ref{B_rho2}) is ensured with
\begin{equation}
C = b_0 \left(b_0^2 - r_0^2 \right)^{\frac{3}{2}} .
\label{C}
\end{equation}
Thus, the wormhole throat constraint also reads $r_{\text{thr}}^2+z_{\text{thr}}^2=b_0^2$, which is 
 equivalently to (\ref{Constraint}). Like the Morris-Thorne case, the wormhole throat extends to the bulk.
 But in this case, it is protected by the cosmological horizon since $\vert y_{\text{thr}} \vert \leq \vert y_{c} \vert $.
 The latter inequality, with the aid of (\ref{Constraint}) and (\ref{y_c}), 
 is valid for
 \begin{equation}
 2b_0^2-\frac{3}{\Lambda_{4d}} \leq r_{\text{thr}}^2 \leq b_0^2.
 \end{equation}
 
The matter content for the five-dimensional spacetime is studied from the field equations (\ref{5d-FE}) like the
previous case.
The energy-momentum tensor components in the $(t,\rho,\theta,\phi,\chi)$ coordinates are
\begin{align}
T^{(B) t}_{\ \ \ \ \  t} = & \ \frac{1}{\kappa_5^2}\Bigg \{ \frac{C}{\left(\rho^2 -r_0^2 \right)^{\frac{5}{2}}}\Bigg[\left(\frac{4\rho^2}{r_0^2}-10+\frac{9r_0^2}{2\rho^2} \right)k \cos \chi \nonumber \\
& -\frac{3r_0^2}{\rho^3}\Bigg] -\Lambda_{5d} \Bigg \}, \\
T^{(B)\rho}_{\ \ \ \ \  \rho}  =  &   \frac{1}{\kappa_5^2}\Bigg \{\frac{C}{\left(\rho^2 -r_0^2 \right)^{\frac{3}{2}}}\Bigg[\frac{4}{r_0^2 \rho} + \frac{3k \cos \chi}{\rho^2} -\frac{3}{\rho^3} \Bigg] -\Lambda_{5d}  \Bigg \}, \\
T^{(B) \theta}_{\ \ \ \ \  \theta} = & \ T^{(B) \phi}_{\ \ \ \ \  \phi}=  T^{(B) \chi}_{\ \ \ \ \  \chi} \nonumber \\
 = & \ \frac{1}{\kappa_5^2}\Bigg \{\frac{C}{\left(\rho^2 -r_0^2 \right)^{\frac{5}{2}}} \Bigg[-\left(\frac{4\rho^2}{r_0^2} - 7 + \frac{9r_0^2}{2\rho^2} \right)k \cos \chi \nonumber \\
 & - 3 \left(\rho -\frac{r_0^2}{2\rho} \right)k^2 \cos^2 \chi \Bigg]  -\Lambda_{5d}  \Bigg \}.
\end{align} 
Thus, like the Morris-Thorne wormhole previously studied, 
the energy-momentum tensor is diagonal in this coordinate system, 
that is, such a tensor is written as  $diag(-\rho_E, p_1, p_2,p_2,p_2)$, 
and the limit of its components is given by Eqs. (\ref{Limit1})-(\ref{Limit2}) like the Morris-Thorne wormhole.

\begin{figure}
\begin{centering}
\includegraphics[trim=0.4cm 0.2cm 1.9cm 0cm, clip=true,scale=0.55]{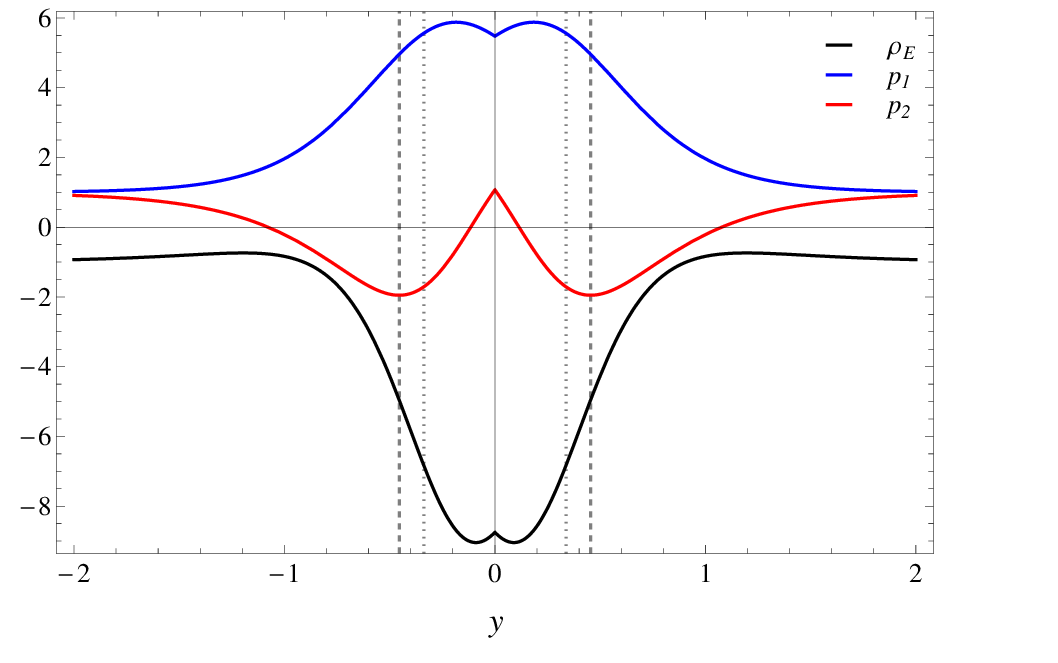}
\par\end{centering}
\caption{Components of the bulk energy-momentum tensor and the weak energy violation close to the brane (located
 at $y=0$)  for the Molina-Neves case. In this graphic, one uses 
 $\kappa_5=\Lambda_{4d}=1$, $\Lambda_{5d}=-1$, $C=5$, $k=1/ \sqrt{2}$, $b_0=1.65$, and the radial coordinate is $r=b_0$. 
The vertical dashed and dotted lines indicate the cosmological horizon 
 and the wormhole throat in the extra dimension, respectively, $y_c=\pm 0.45$ and $y_{\text{trh}}=\pm 0.3$.}
\label{Energy-MN}
\end{figure}

In order to see the dependence of the energy density and pressures on the extra coordinate $y$, the coordinates
$(t,r,\theta,\phi,y)$ should be adopted. In this coordinate system, one has
\begin{align}
\rho_{E} = & \frac{1}{\kappa_5^2} \Bigg \{ \Lambda_{5d} + \frac{C}{\Delta ^{\frac{5}{2}}}\Bigg[ \frac{k(e^{k\vert y \vert}-1)(4\Delta -6)}{r_0^5  [(e^{k \vert y \vert}-1)^2+k^2 r^2]^{\frac{1}{2}} } \nonumber \\ 
& + \frac{3k^3(3e^{k\vert y \vert}-4)}{2 r_0^3  [(e^{k \vert y \vert}-1)^2+k^2 r^2]^{\frac{3}{2}} } \Bigg] \Bigg \}, \label{rho-MN} \\
p_1 = & \frac{1}{\kappa_5^2} \Bigg \{- \Lambda_{5d} + \frac{C}{\Delta ^{\frac{3}{2}}}\Bigg[ \frac{k^3(4\Delta +3e^{\vert y \vert} -2)}{r_0^3  [(e^{k \vert y \vert}-1)^2+k^2 r^2]^{\frac{3}{2}} } \Bigg]  \Bigg \},  \label{p1-MN} \\
p_2 = &  \frac{1}{\kappa_5^2} \Bigg \{ - \Lambda_{5d}  + \frac{C (e^{k \vert y \vert}-1)}{\Delta ^{\frac{5}{2}}}\Bigg[ \frac{k(6-4\Delta -3e^{k\vert y \vert})}{r_0^5  [(e^{k \vert y \vert}-1)^2+k^2 r^2]^{\frac{1}{2}} } \nonumber \\ 
& + \frac{3k^3(e^{k\vert y \vert}-4)}{2 r_0^3  [(e^{k \vert y \vert}-1)^2+k^2 r^2]^{\frac{3}{2}} } \Bigg]  \Bigg \},\label{p2-MN}
\end{align}
with the dimensionless parameter $\Delta$ defined as
\begin{equation}
\Delta = \frac{(e^{k \vert y \vert}-1)^2 + k^2 (r^2-r_0^2)}{k^2 r_0^2}.
\end{equation}
Once again, the five-dimensional fluid is anisotropic, and the weak condition is violated 
according to Fig. \ref{Energy-MN}.
In the $(t,r,\theta,\phi,y)$ coordinates, $T^{(B)M}_{\ \ \ \ \ N}$ also has two off-diagonal 
components $T^{(B)r}_{\ \ \ \ \  y}$ and $T^{(B)y}_{\ \ \ \ \ r}$, which are zero for the limits 
$y\rightarrow 0$ and $y \rightarrow \infty$.
Therefore, the five-dimensional anti-de Sitter spacetime is recovered in the latter limit. 
Using the coordinates $r$ and $y$, it is straightforward to verify the weak energy violation. Close to the brane, that is,
for small values of $y$, one has
\begin{equation}
\rho_{E}+ p_2 \simeq - \frac{3C}{\kappa_5^2 (r^2-r_0^2)^{\frac{5}{2}}}\left(\frac{k \vert y \vert}{r} + \frac{r_0^2}{2r^3}\right) <0.
\end{equation}

Let us assume that (\ref{A-MN}) and (\ref{B-MN}) are the bulk metric functions 
and the Molina-Neves wormhole (\ref{MN}) is an induced metric on
the brane. Thus, the \enquote{electric} part of the Weyl tensor is written as
\begin{align}
E_{\mu\nu}  \bigg \vert_{y \rightarrow 0 } = \  &  \frac{1}{r\left(r^2 -r_0^2 \right)^{\frac{3}{2}}-C} \nonumber \\
& \times  \left(\begin{array}{cccc}
\mathcal{E}_{t} h_{tt}\\
 & \mathcal{E}_{r}  h_{rr} \\
 &  &  \mathcal{E}_{\theta}  h_{\theta \theta} \\
 &  &  & \mathcal{E}_{\phi}  h_{\phi \phi}
\end{array}\right),
\label{E-MN}
\end{align}
where
\begin{align}
\mathcal{E}_t =  & \ \frac{r  \left(r^2- r_0^2\right)^{\frac{3}{2}}}{6 r^2-9 r_0^2} -\frac{C \left(4 r^2-3 r_0^2\right)}{6 r^2 \left(2 r^2-3 r_0^2\right)} \nonumber \\
&  +\frac{C^2 \left(2 r^2- r_0^2 \right)}{4 r^3 \left(r^2- r_0^2\right)^{\frac{5}{2}}}, \\
\mathcal{E}_r = & - \frac{r  \left(r^2- r_0^2\right)^{\frac{3}{2}}}{2 r^2 -3 r_0^2}- \frac{C \left(4 r^4 - 9 r_0^2 r^2  + 3 r_0^4\right)}{2 r^2 \left(2 r^2-3 r_0^2\right) \left(r^2- r_0^2\right)} \nonumber \\
&  +\frac{C^2 \left(2 r^2 -  r_0^2\right)}{4 r^3 \left(r^2- r_0^2\right)^{\frac{5}{2}}}, \\
\mathcal{E}_{\theta} = & \ \mathcal{E}_{\phi} =  \frac{r  \left(r^2- r_0^2\right)^{\frac{3}{2}}}{6 r^2-9 r_0^2} +\frac{C \left(8 r^4-17 r_0^2 r^2 +6 r_0^4\right)}{6 r^2 \left(2 r^2-3 r_0^2\right) \left(r^2- r_0^2\right)} \nonumber \\
& -\frac{C^2 \left(2 r^2- r_0^2\right)}{4 r^3 (r^2- r_0^2)^\frac{5}{2}}.
\end{align}
It is worth emphasizing two points: firstly, the condition $E^{\mu}_{\ \mu}=0$ is satisfied. Secondly,
the tensor $E_{\mu\nu}$ is singular for $r=r_c=\sqrt{3/2}\ r_0$ and at the throat, $r=b_0$, 
where Eq. (\ref{C}) is valid. This problem will be circumvented when the effective field equations were written.

Before writing the effective field equations we need to calculate the tensor (\ref{T_eff}). Assuming again that 
(\ref{A-MN}) and (\ref{B-MN}) are the bulk metric functions 
and the metric (\ref{MN}) is an induced metric on the brane, one has
\begin{align}
T_{\mu\nu}^{(\text{eff})} = & \ \frac{(r^2-r_0^2)^{\frac{3}{2}}}{\kappa_5^2 k [r\left(r^2 -r_0^2 \right)^{\frac{3}{2}}-C]} \Bigg[3k^2 r h_{\mu\nu} \nonumber \\
& + \left(\begin{array}{cccc}
\mathcal{T}_t h_{tt}\\
 &\mathcal{T}_r  h_{rr} \\
 &  & \mathcal{T}_{\theta} h_{\theta \theta} \\
 &  &  & \mathcal{T}_{\phi} h_{\phi \phi}
\end{array}\right) \Bigg],
\label{T-MN}
\end{align}
in which
\begin{widetext} 
\begin{align}
\mathcal{T}_t = & - \frac{12 r^3 - 19 r_0^2 r}{3 r_0^2 \left( 2 r^2 -3 r_0^2\right)} 
+ \frac{C \left[\left(16-36 k^2 r_0^2\right)r^6  -  \left(32 - 90 k^2 r_0^2 \right)r_0^2 r^4 + \left(13-54 k^2 r_0^2\right) r_0^4r^2 - 3 r_0^6\right]}{6 r_0^2 r^2  \left(2 r^2-3 r_0^2\right) (r^2- r_0^2)^{\frac{5}{2}}} \nonumber \\
& + \frac{C^2 \left(8 r^4-14 r^2 r_0^2-3 r_0^4\right)}{12 r_0^2 r^3  \left(r^2 - r_0^2\right)^4}, \\
\mathcal{T}_r = & -\frac{4 r^3 - 5 r_0^2 r}{r_0^2 \left(2 r^2 - 3 r_0^2\right)} + \frac{C \left[ \left(16 -12 k^2 r_0^2 \right)r^6 - \left(48 -30 k^2 r_0^2\right)r_0^2 r^4 + \left(43 - 18 k^2 r_0^2\right) r_0^4r^2 - 9 r_0^6\right]}{2 r_0^2 r^2 \left(2 r^2-3 r_0^2\right) \left(r^2 - r_0^2\right)^{\frac{5}{2}}}  \nonumber \\
 & - \frac{C^2 \left(8 r^4-14  r_0^2 r^2 + 5 r_0^4\right)}{4 r_0^2 r^3 \left(r^2- r_0^2\right)^4}, \\
 \mathcal{T}_{\theta} = &  \mathcal{T}_{\phi} =  -\frac{12 r^3 - 19 r_0^2 r}{3 r_0^2 \left(2r^2 - 3r_0^2 \right)} +\frac{C \left[ \left(16-36 k^2 r_0^2\right)r^4 - \left(16 -54 k^2 r_0^2 \right)r_0^2 r^2 - 15 r_0^4\right]}{6 r_0^2 r^2   \left(2 r^2-3 r_0^2\right)(r^2-r_0^2)^{\frac{3}{2}}}  + \frac{C^2 \left(8 r^4-14 r^2 r_0^2+9 r_0^4\right)}{12  r_0^2 r^3 \left(r^2-r_0^2\right)^4}.
\end{align}
\end{widetext} 
As we can see, the tensor $T_{\mu\nu}^{(\text{eff})}$ is singular 
at the wormhole throat as Eq. (\ref{C}) is satisfied and at the cosmological horizon $r_c$ 
like the tensor $E_{\mu\nu}$. However, inserting (\ref{E-MN}) and (\ref{T-MN}) 
in Eq. (\ref{Field_eq2}), 
we can see that the Einstein tensor on the brane is the same obtained by Molina and Neves \cite{MN}. Accordingly, 
\begin{equation}
G_{\mu\nu} = -  \left(\begin{array}{cccc}
  \mathcal{G}_t h_{tt}\\
 & \mathcal{G}_r h_{rr}  \\
 &  &  \mathcal{G}_{\theta} h_{\theta \theta}  \\
 &  &  &  \mathcal{G}_{\phi}  h_{\phi \phi} 
\end{array}\right),
\label{G-MN}
\end{equation}
with
\begin{align}
\mathcal{G}_t  = &  \frac{2}{r_0^2} + \frac{C \left(2 r^2-5 r_0^2\right)}{3 r_0^2 r \left(r^2-r_0^2\right)^{\frac{5}{2}}}, \label{G1-MN} \\
\mathcal{G}_r  = & \frac{2}{r_0^2} - \frac{C \left(2r^2 - r_0^2\right)}{r_0^2 r^3 \left(r^2-r_0^2\right)^{\frac{3}{2}}}, \label{G2-MN} 
\end{align}
\begin{equation}
\mathcal{G}_{\theta} =   \mathcal{G}_{\phi} =  \frac{2}{r_0^2} +\frac{C \left(4 r^4-4 r_0^2 r^2 +3 r_0^4\right)}{6 r_0^2 r^3 \left(r^2-r_0^2\right)^{\frac{5}{2}}}.\label{G3-MN} 
\end{equation}
Interestingly, $G_{\mu\nu}$ is regular on the brane for the interval $b_0 \leq r \leq r_c$.
The singularity issues for $E_{\mu\nu}$ and $T_{\mu\nu}^{(\text{eff})}$ are avoided as we write the effective
field equations on the brane. Therefore, the Molina-Neves wormhole (\ref{MN}) could be think of as 
an induced spacetime metric in the context adopted here. That is, from the bulk perspective,
it is a five-dimensional wormhole supported by an exotic fluid described by Eqs. (\ref{rho-MN})-(\ref{p2-MN}). 
From the brane perspective, it is a four-dimensional wormhole on the brane without particles or 
fields that  live exclusively on the brane. 

It is worth noting that Molina and Neves obtained the metric (\ref{MN}) without the bulk metric. Their effective
field equations are $G_{\mu\nu}= - \Lambda_{4d}g_{\mu\nu} -E_{\mu\nu}$, that is, the brane in that case
is asymptotically de Sitter due to the four-dimensional cosmological constant. 
Regarding the field equations adopted by the authors, the Einstein tensor expression (\ref{G-MN}) indicates two points:
 the first term on the right side of Eqs. (\ref{G1-MN})-(\ref{G3-MN}) is the cosmological constant term, 
and the second term stands for the traceless tensor $E_{\mu\nu}$, which is, according to  Eq. (\ref{E-MN}), very 
different from the obtained here. But, as we saw, the effective field equation adopted in this article
is Eq. (\ref{Field_eq2}), thus the exotic fluid in the bulk supports both an asymptotic anti-de Sitter behavior 
in the bulk and the asymptotic de Sitter behavior on the brane. Consequently, $\Lambda_{5d}$ and $\Lambda_{4d}$ 
are just \enquote{effective} cosmological constants in the approach adopted here.  

\section{Final remarks}
\label{Sec-V}
Historically, metaphysics is the area of knowledge that interprets the origin of the physical world assuming
a world or something beyond the phenomena, whether the Platonic world of Forms \cite{Plato} or the 
Kantian thing in itself \cite{Kant}. In today's physics, extra dimensions could be source or
origin of phenomena as well. But in this case, extra dimensions would have the same ontological status of the 
four-dimensional physical world.

In the brane-world context adopted in this work---in which our four-dimensional universe would be embedded in a 
five-dimensional spacetime---two five-dimensional wormhole metrics were obtained.
Following the Randall-Sundrum II model, the bulk or the five-dimensional spacetime is asymptotically
anti-de Sitter, and the four dimensional brane (our universe) could be either asymptotically flat or de Sitter.
We saw that spacetime metrics like the Morris-Thorne wormhole and the Molina-Neves wormhole are viable on the brane.
Such wormholes are induced on the brane due to the exotic fluid that supports the five-dimensional spacetime.
Interestingly, there is no need to add particles or fields on the brane. The gravitational effect on the brane 
is entirely a consequence of the bulk influence on the brane.

When faced with data of unknown entities like dark matter and dark energy, 
a brane-world context could potentially justify 
such phenomena without particles or fields on the brane. These phenomena would be effects of 
an extra dimension or a world beyond, like the \enquote{effective} four-dimensional 
cosmological constant obtained here. 

\section*{Acknowledgments}
I would like to thank Conselho Nacional de Desenvolvimento Científico e Tecnológico (CNPq), Brazil, 
(Grant No. 170579/2023-9) for the financial support and the ICT-Unifal for the kind hospitality. 
Also, I thank an anonymous referee for valuable comments.


\end{document}